\documentclass{INTERSPEECH2023}

\interspeechcameraready

\usepackage{multirow}
\usepackage{url}
\title{Automatic Speech Disentanglement for Voice Conversion using \\ Rank Module and Speech Augmentation}
\name{Zhonghua Liu$^1$, Shijun Wang$^2$, Ning Chen${^{1,*}}$\thanks{*Corresponding author}}
\address{
  $^1$East China University of Science and Technology, Shanghai, China\\
  $^2$University of St.Gallen, St.Gallen, Switzerland}
\email{y80210144@mail.ecust.edu.cn, shijun.wang@unisg.ch, nchen@ecust.edu.cn}

\begin{document}

\maketitle
 
\begin{abstract}
Voice Conversion (VC) converts the voice of a source speech to that of a target while maintaining the source's content. 
Speech can be mainly decomposed into four components: content, timbre, rhythm and pitch.
Unfortunately, most related works only take into account content and timbre, 
which results in less natural speech. 
Some recent works are able to disentangle speech into several components, but they require laborious bottleneck tuning or various hand-crafted features, each assumed to contain disentangled speech information.
In this paper, we propose a VC model that can automatically disentangle speech into four components using only two augmentation functions, without the requirement of multiple hand-crafted features or laborious bottleneck tuning. 
The proposed model is straightforward yet efficient, and
the empirical results demonstrate that our model can achieve a better performance than the baseline, regarding disentanglement effectiveness and speech naturalness.

\end{abstract}

\section{Introduction}

Voice Conversion (VC) converts one’s voice to the voice of a target speaker while preserving the linguistic content of the source speech \cite{lorenzo2018voice, sisman2020overview}.
VC brings benefits for many interesting and mature applications, such as the famous movie dubbing, AI anchor, intelligent voice conversion navigation, etc.
In recent years, with the development of deep neural networks \cite{Taylor2016AudiotoVisualSC, lo2019mosnet, jia2019direct}, voice conversion methods are also gradually improving and can reach excellent VC performance.

At present, to achieve successful VC, some approaches \cite{Huang2021OnPM, Liu2019UnsupervisedEL, Huang2020TheSB} use auxiliary models like Automatic Speech Recognition (ASR) or Text-To-Speech (TTS) models to achieve VC. In \cite{Fang2018HighQualityNV, Kaneko2018CycleGANVCNV, Kameoka2018StarGANVCNM, Li2021StarGANv2VCAD}, the authors employ Generative Adversarial Networks (GANs) to teach the decoder to generate speech that sounds like the target speakers. However, GAN-based models are usually hard to train.
Disentanglement-based approaches such as \cite{Wu2020VQVCOV, Xiao2022DGCVectorAN, Wang2021OneShotVC, Wang2021VQMIVCVQ, Luong2021ManytoManyVC} aim to split the speech into spoken content and speaker characteristic (i.e. timbre). 
These models have been widely studied because they can achieve excellent performance without using auxiliary models like ASR and it is easier to train compared with GAN-based VC models.

% At present, the popular voice conversion methods include using Automatic Speech Recognition and Text-To-Speech method, Generative Adversarial Network method, and disentanglement-based method.
% ASR-TTS-based method such as in \cite{Huang2021OnPM,Liu2019UnsupervisedEL,Huang2020TheSB}, they use complex TTS models such as Tacotron2 and FastSpeed to realize voice conversion.
% GAN-based method such as in \cite{Fang2018HighQualityNV,Kaneko2018CycleGANVCNV,Kameoka2018StarGANVCNM,Li2021StarGANv2VCAD} they use they use discriminator that teaches the decoder to generate speech that sounds like the target speaker,which are hard to train.
% Disentangled-based method split the speech into spoken content and speaker characteristic such as \cite{Wu2020VQVCOV,Xiao2022DGCVectorAN,Wang2021OneShotVC,Wang2021VQMIVCVQ,Luong2021ManytoManyVC} have been widely studied because of it can achieve excellent performance without using other models like ASR. And it is easier to train compared with GAN-based VC models.

However, most disentanglement-based approaches neglect the fact that speech carries rhythm and pitch \cite{Gan2022IQDUBBINGPM} as well as content and timbre. 
Content is the linguistic information conveyed by the speaker; rhythm indicates how quickly the speaker speaks; pitch is the perceived “highness” or “lowness” of a voice; timbre is perceived as the voice characteristic.
All of these critical components carry important information and are entangled in the speech. 
If models simply consider timbre and ignore other components, the generated speech might be less natural and expressive.
As a result, decoupling these four key components is crucial for successful VC.
% However, most of the related approaches only take into account timbre \cite{Lin2020FragmentvcAV,Xiao2022DGCVectorAN,Wang2021OneShotVC,Luong2021ManytoManyVC} while ignoring other components, which might decrease the performance when trying to generate speech that is both natural and expressive. Therefore, decoupling the rhythm and pitch from the speech is crucial for VC.

SpeechSplit \cite{Qian2020UnsupervisedSD} and SpeechSplit2.0 \cite{Chan2022SpeechSplit20US} are two approaches that attempt to decouple the aforementioned speech components.
Both models have a similar architecture, with content, rhythm, pitch encoders and one decoder.
For SpeechSplit, the authors feed each encoder different types of hand-crafted features, that are manually designed to contain individual disentangled speech information.
% SpeechSplit disentangles speech by applying signal processing techniques like rhythm resampling or pitch normalization.
% % However, techniques like pitch normalization might severely damage the pitch information, causing pitch conversion imperceptible.
However, careful bottleneck tuning (usually utilizing small bottlenecks) is required for successful disentanglement, which sacrifices the quality of the generated speech.
SpeechSplit2.0 is built upon SpeechSplit. 
By employing more signal processing techniques,
hand-crafted features that are fed to different encoders can hold more distinct speech information.
This strategy allows SpeechSplit2.0 to disentangle the speech without laborious bottleneck tuning. 

However, the effectiveness of disentanglement in SpeechSplit2.0 highly depends on the hand-crafted features.
It would be preferable if the model is able to automatically disentangle the speech, as it could save time on manually selecting the features. 
Moreover, an automatic disentanglement process can potentially produce disentangled representations that are more effective and accurate, 
since it can reduce bias and subjectivity that could be introduced by manual selection of features.

% Nevertheless, these two models require numerous hand-crafted features and signal processing techniques, which might not be straightforward to implement.
% Additionally, there is still room for improvement regarding speech naturalness.
% \textcolor{red}{otherwise, the rhythm encoder can encode all four aspects.}
% SpeechSplit disentangles speech into content, rhythm, pitch, and timbre using
% three encoders with carefully tuned bottlenecks.
% However, improper bottleneck design will lead to poor conversion results. Small bottlenecks might degrade the speech quality because small bottlenecks are not able to carry enough information to achieve satisfactory speech generation. 
%  Furthermore, it is also time-consuming and less robust for some converted samples.
% SpeechSplit2.0 solves these issues by applying signal processing techniques to alleviate the laborious bottleneck tuning, but the pitch smoother introduces artifacts in the smoothed audio mainly due to pitch tracking errors, which ultimately affect the naturalness. 
 
In this paper, we propose a VC approach,  
whose architecture is similar to SpeechSplit or SpeechSplit2.0. 
But instead of feeding numerous hand-crafted features, our model can automatically disentangle the speech with only two augmentation transformations (pitch modification and rhythm adjustment), while bottleneck tuning is not required.
The idea is inspired by \cite{Souri2015DeepRA, Wang2021ZeroshotVC}, where the authors successfully used the Rank module to enable their models to automatically extract effective disentangled representations from the data. Thereby, it is reasonable to consider using the Rank module for disentangling speech as well.
Precisely, we force the model to disentangle speech components by ranking the speech and its augmented version.
For example, if we increase the pitch, then given the pitch representations from the pitch encoder,
the model should rank the augmented version higher than the original version. 
At the same time, as the rhythm remains the same after the augmentation, the rhythm information from both versions should be ranked equally.
Furthermore, we apply the idea from SimCLR \cite{Chen2020ASF} to effectively extract invariant representations from the data, i.e. content information, since content information is independent of the changes in pitch, timbre or rhythm.
Particularly,
we force the model to attract content representations from the original and the augmented versions, while repelling content representations from all other speech data in the mini-batch.
% The proposed work can (1) reach a better performance than SpeechSplit1\&2 without careful bottleneck tunning. (2) better robustness on different datasets and tasks(3) The results of our method's pitch conversion experiments demonstrate a more natural-sounding and improved pitch contour. (4) unlike SpeechSplit1\&2 that need variable signal processing technics, our work relies only on augmentation.(5)unlike SpeechSplit1\&2 use Wavnet vocoder,the conversion speed is very slow,We replaced it with HiFiGAN vocoder with faster conversion speed.

We summarize our contributions as: 
(1) we propose a voice conversion model that is able to disentangle the speech into content, timbre, rhythm and pitch components;
(2) the disentanglement process is automatic, and it does not require bottleneck fine-tuning or various hand-crafted features;
(3) our empirical results show that, compared to the state-of-the-art SpeechSplit2.0, the proposed model achieves better disentanglement performance and generates more natural speech.

% (1) reach better performance then SpeechSplit1\&2 without careful bottleneck tuning. (2) better robustness (3) better pitch conversion (4) unlike SpeechSplit1\&2 that need variable signal processing technologies, our work relies only on augmentation. (5)Use vocoder with faster conversion speed：HiFi-GAN

\section{Methodology}

We train our model in two steps. First, we train multiple encoders to extract content, rhythm and pitch information from the speech. 
For the second step, we train the decoder with the timbre information and other extracted speech components.

\subsection{Speech Encoders}
\label{sec:encoder}

\begin{figure}[t]
  \centering
  \includegraphics[width=0.9\linewidth]{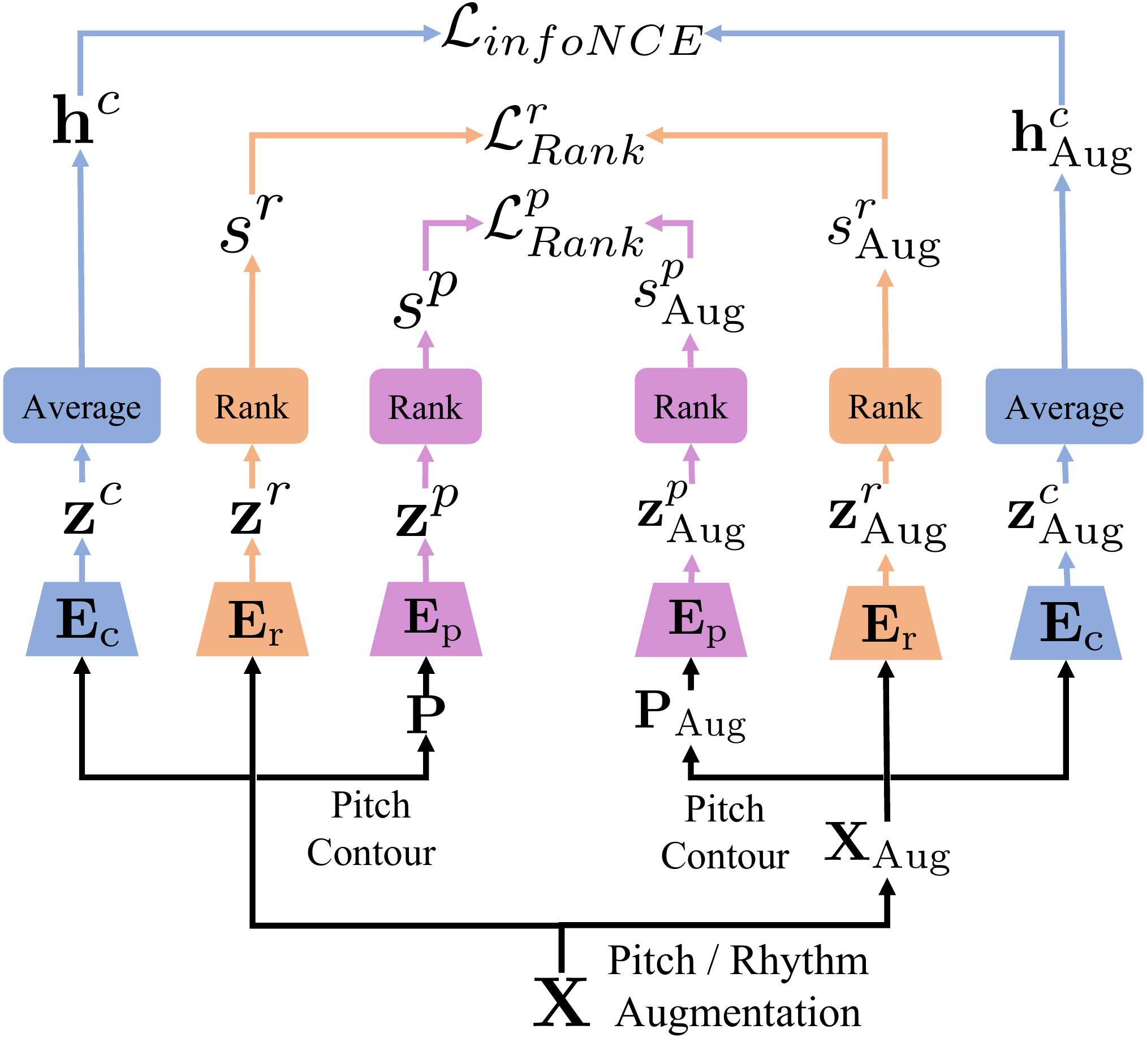}
  \caption{Speech $\mathbf{X}$ and its augmented version $\mathbf{X}_{\text{Aug}}$ are sent to our encoders to produce pairs ($\textbf{z}^c$, $\textbf{z}_{\text{Aug}}^c$), ($\textbf{z}^r$, $\textbf{z}_{\text{Aug}}^r$), ($\textbf{z}^p$, $\textbf{z}_{\text{Aug}}^p$) for content, rhythm and pitch, respectively. After the Rank module, we rank the rhythm and pitch information between the original and the augmented speech with the losses $\mathcal{L}_{rank}^r$ and $\mathcal{L}_{rank}^p$. Meanwhile, we apply $\mathcal{L}_{infoNCE}$ to attract the content representations from the original and the augmented versions, while repelling other content representations in this mini-batch.}
  \label{rank_fig}
  \vspace{-0.3cm}
\end{figure}

Like SpeechSplit2.0, to extract content, rhythm and pitch representations from the speech, we employ three encoders in our model.
The training pipeline is shown in Fig.~\ref{rank_fig}. 

As we can see, we first augment the speech data $\mathbf{X}$ into $\mathbf{X}_{\text{Aug}}$. 
In our work, we use \textit{SoX}\footnote{\url{https://sox.sourceforge.net/}} to apply two augmentation functions $\texttt{PitchAug}(\mathbf{X}, \tau^p)$ and $\texttt{RhythmAug}(\mathbf{X}, \tau^r)$ to modify the speech rhythm and pitch, where hyperparameter $\tau \in (0, 1)$ indicates the augmentation intensity. Given a specific augmentation function, $\tau < 0.5$ means negative augmentations (decrease pitch/rhythm), while $\tau > 0.5$ means positive augmentations (increase pitch/rhythm). Lastly, $\tau = 0.5$, which indicates no augmentation is applied.
Please keep in mind that during the training, each sample is augmented only once by a randomly selected augmentation function.

We then send speech $\mathbf{X}$ and its augmentation version $\mathbf{X}_{\text{Aug}}$ into our encoders. On the left, $\mathbf{X}$ is fed into content encoder $\mathbf{E_c}$ and rhythm encoder $\mathbf{E_r}$, while its pitch contour $\mathbf{P}$ is fed into pitch encoder $\mathbf{E_p}$. 
With these three encoders, we can get representations $\mathbf{z}^c$, $\mathbf{z}^r$ and $\mathbf{z}^p$. Each one is a sequence and represents content, rhythm, and pitch information, respectively.
Afterward, we apply two Rank modules (linear layers) to map $\mathbf{z}^r$ and $\mathbf{z}^p$ into two individual scores $s^r$ and $s^p$. 
Both scores are scales, $s^r$ indicates how fast the speech is, while $s^p$ indicates how high the pitch of the speech is.
Meanwhile, the content representation $\mathbf{z}^c$ is first averaged into a vector $\mathbf{h}^c$ for further operations, this is inspired by SimCLR \cite{Chen2020ASF}, where the authors use projection heads to simplify the high-dimensional image representations and make them easier to compare with other image representations. 
Similarly, as we can see from the right path, with the augmented version $\mathbf{X}_{\text{Aug}}$, we can get representations $\mathbf{z}_{\text{Aug}}^c$, $\mathbf{z}_{\text{Aug}}^r$ and $\mathbf{z}_{\text{Aug}}^p$; two scores $s_{\text{Aug}}^r$ and $s_{\text{Aug}}^p$; one vector $\mathbf{h}_{\text{Aug}}^c$.

In order to force the encoders $\mathbf{E_c}$, $\mathbf{E_r}$ and $\mathbf{E_p}$ to produce disentangled representations. We first feed the rhythm and pitch score differences into a Sigmoid function:
\begin{equation}
\begin{gathered}
d^r = \frac{1}{1+e^{-(s^r-s_{\text{Aug}}^r)}}, \\
d^p = \frac{1}{1+e^{-(s^p-s_{\text{Aug}}^p)}},
\end{gathered}
\end{equation}
then we apply the rank loss on $d^p$ and $d^r$:
\begin{equation}
\begin{gathered}
\mathcal{L}_{rank}^r = -\tau^{r} \text{log}(d^{r}) - (1-\tau^{r}) \text{log}(1-d^{r}), \\
\mathcal{L}_{rank}^p = -\tau^{p} \text{log}(d^{p}) - (1-\tau^{p}) \text{log}(1-d^{p}),
\label{loss:rank}
\end{gathered}
\end{equation}
as we mentioned before, $\tau^r$ and $\tau^{p}$ are two hyperparameters indicating the augmentation intensity. 
And our target is to force our encoders to produce disentangled representations by recognizing this augmentation intensity.

As an example, if we perform augmentation $\texttt{PitchAug}$ and increase the pitch ($\tau^p > 0.5$) of $\mathbf{X}$. Then, to decrease the pitch Rank loss $\mathcal{L}_{rank}^p$, the model needs to assign a higher score to $s_{\text{Aug}}^p$ than $s^p$, which means the pitch representation $\textbf{z}^p$ from the pitch encoder $\mathbf{E_p}$ is encouraged to carry effective pitch information.
At the same time, since rhythm information is not modified by augmentation $\texttt{PitchAug}$, $\tau^r$ is set to 0.5. 
Therefore, the model has to assign the same score to both $s^r$ and $s_{\text{Aug}}^r$ to decrease the loss $\mathcal{L}_{rank}^r$. 
In other words, the rhythm encoder $\mathbf{E_r}$ is forced to be insensitive to the change of pitch, i.e., rhythm representation $\textbf{z}^r$ and pitch representation $\textbf{z}^p$ are disentangled from each other.

Furthermore, to make sure the content encoder $\mathbf{E_c}$ only produces the content-related representations, we apply the infoNCE loss from \cite{Chen2020ASF,Oord2018RepresentationLW} on $\textbf{h}^c$ and $\textbf{h}_{\text{Aug}}^c$, we make a slightly rephrased to Equation (1) from [23]:
\begin{equation}
\label{loss:infonce}
\mathcal{L}_{infoNCE} = -\text{log} \frac{\text{sim}(\textbf{h}^c, \textbf{h}_\text{Aug}^c)} {\text{sim}(\textbf{h}^c, \textbf{h}_\text{Aug}^c) + \sum\limits_{\mathbf{X}_{\text{Neg}}}\text{sim}(\textbf{h}^c, \textbf{h}_{\text{Neg}}^c) }, \\
\end{equation}
where sim($\cdot, \cdot$) is the exponential dot product of the two inputs, with a temperature parameter $t$. $\mathbf{X}_{\text{Neg}}$ denotes all speech samples in a mini-batch except $\mathbf{X}$, while $h_{\text{Neg}}^c$ denotes the content vectors derived from $\mathbf{X}_{\text{Neg}}$.

With $\mathcal{L}_{infoNCE}$, we force the content representations from the same speech to be similar, regardless of which augmentation methods we use. 
Meanwhile, content representations derived from different speech samples are forced to be dissimilar. 
Such loss enables the content encoder $\mathbf{E_c}$ to be only focused on the invariant information of the speech, i.e. content information.

Finally, we train the three encoders with the loss:
\begin{equation}
\mathcal{L}_{encoder} = \mathcal{L}_{rank}^r + \mathcal{L}_{rank}^p  + \mathcal{L}_{infoNCE}.
\end{equation}

\subsection{Speech Decoder}
\label{sec:vc}

\begin{figure}[t]
  \centering
  \includegraphics[width=0.7\linewidth]{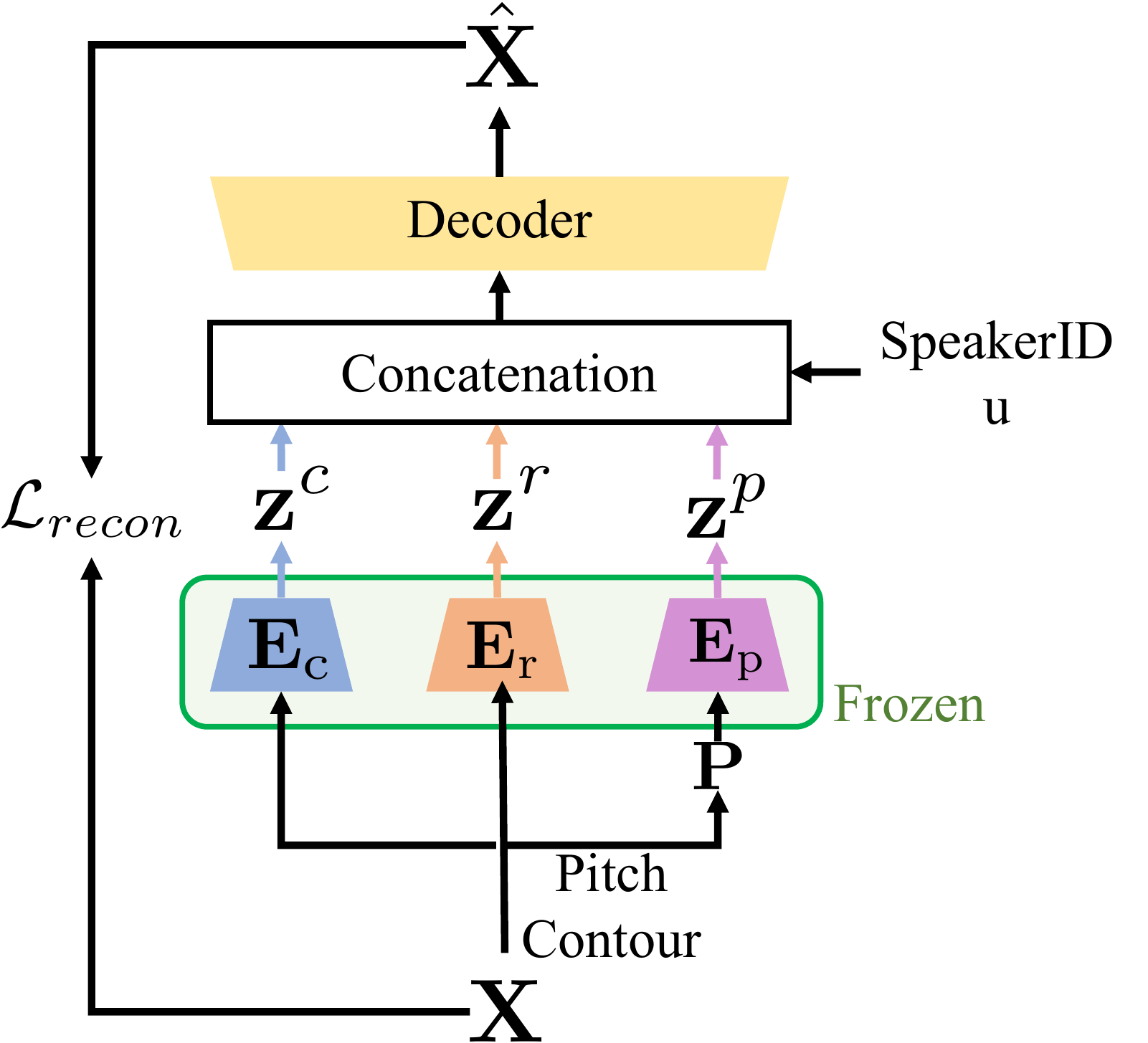}
  \caption{The training of the voice conversion.}
  \label{recon_fig}
  \vspace{-0.3cm}
\end{figure}

For VC, we need to provide timbre information to represent speaker characteristics. We follow SpeechSplit2.0 and use speaker ID as timbre information. 
Then, we need to train a decoder to perform voice conversion.
The training pipeline is shown in Fig.~\ref{recon_fig}. 
From the input speech $\mathbf{X}$, the content, rhythm and pitch representations are extracted by pre-trained and frozen encoders.
Meanwhile, speaker ID $u$ is converted to speaker embedding.
The rest is similar to SpeechSplit2.0, all these representations are concatenated and fed to the decoder. 
And the decoder generates $\mathbf{\hat{X}}$, and we apply the reconstruction loss (mean square error) on $\mathbf{X}$ and $\mathbf{\hat{X}}$:
\begin{equation}
\mathcal{L}_{recon} = ||\mathbf{X} - \mathbf{\hat{X}}||^2.
\end{equation}

\subsection{Implementation}
Our model's architecture is the same as SpeechSplit2.0 \cite{Chan2022SpeechSplit20US}. But rather than training all of the modules at once, we train our model in two steps, because we found it can achieve better performance. 
In the first step (Sec.~\ref{sec:encoder}), we train our encoders for 30k iterations, and set the learning rate to 1e-6, while the temperature $t$ in Eq.~\ref{loss:infonce} is 0.1. For the second step (Sec.~\ref{sec:vc}), The decoder is trained using a 1e-4 learning rate over 600k iterations. Adam optimizer is used for both steps. Demo audio samples can be found at \url{https://hhhuazi.github.io/}

\section{Experiments}

\subsection{Experiment setup}

\noindent 
\textbf{Dataset and Data Preprocessing}: We use VCTK corpus \cite{Veaux2017CSTRVC} to train and evaluate our models, which is a popular dataset for VC tasks. VCTK includes 109 English speakers, each speaker reads about 400 sentences. All audio recordings are down-sampled to 16kHz. And we extract Mel-Spectrograms from waveform files by using a 50-millisecond window and a 50 percent overlap ratio, with 80 Mel Coefficients.

\noindent 
\textbf{Baseline}: We employ SpeechSplit2.0 \cite{Chan2022SpeechSplit20US} as our baseline. 
SpeechSplit2.0 is a state-of-the-art VC model that is able to disentangle speech like the proposed model.

\noindent 
\textbf{Evaluational Setup}: We use HiFi-GAN \cite{Kong2020HiFiGANGA} to convert generated Mel-Spectrograms to waveforms. To perform objective and subjective tests, we randomly select 20 unseen source and target pairs to generate synthesized samples from both models.
20 subjects participate in the subjective evaluations.

% The proposed VC network is trained using the ADAM optimizer (learning rate is e-4, $\beta_{1}$ = 0.9, $\beta_{2}$ = 0.98) with a batch size of 16 for 800k steps. The model structure of content encoder, prosody encoder, pitch encoder and decoder are the same as SpeechSplit. It is worth mentioning that We replaced the WaveNet vocoder \cite{Oord2016WaveNetAG} , which mainly uses the expansion causal convolutional network to improve the synthesis effect but will lead to its large amount of computation and can not achieve real-time synthesis with the HiFi-GAN \cite{Kong2020HiFiGANGA} vocoder, which proposed a residual structure, alternately use holed convolution and ordinary convolution to increase the receptive field to ensure the synthetic sound quality and improve the reasoning speed.

\begin{figure*}
  \centering
  \includegraphics[width=\textwidth]{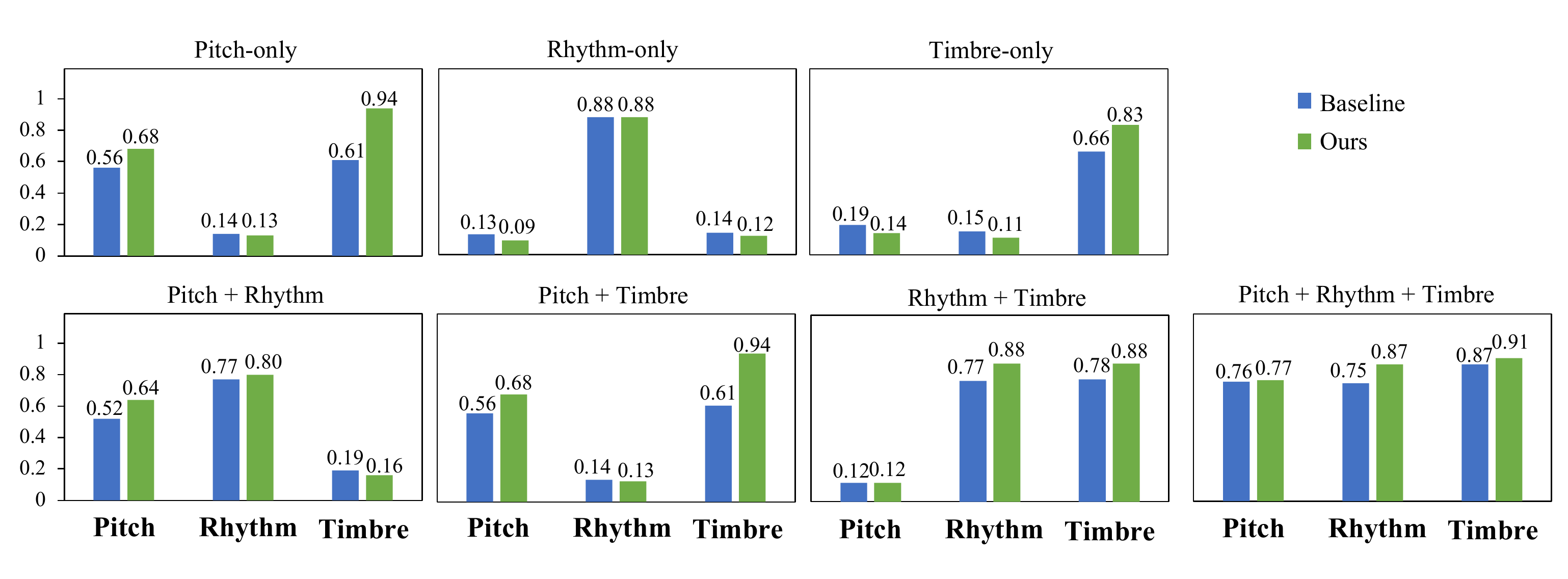}
  \caption{Conversion rate}
  \label{conversion_fig}
  \vspace{-0.4cm}
\end{figure*}

\subsection{Conversion Rate}

To evaluate the disentanglement effect, we follow \cite{Chan2022SpeechSplit20US} and evaluate the conversion rate of the baseline and our model. 
In this subjective test, each subject is first asked to listen to the source and target reference speech in random order.
Then, by listening to a synthesized sample, the subject selects which reference speech has more similar components (pitch, rhythm, or timbre) to the synthesized sample. 
Afterward, we can compute the conversion rate, defined as the percentage of answers selecting the target reference.
The conversion rate can reflect the effectiveness of the decoupling.
For example, if we only convert the timbre, then the timbre conversion rate is supposed to be high, while the pitch and rhythm conversion rates should be low. In our experiments, we test the conversion rate on all 7 combinations (Pitch-only, Rhythm-only, Timbre-only, Pitch+Rhythm, Pitch+Timbre, Rhythm+Timbre, Pitch+Rhythm+Timbre). 
The results are shown in Fig.~\ref{conversion_fig}.

As we can see, although both models have the ability to disentangle the speech, our model significantly outperforms the baseline for almost all combinations. 
Such results demonstrate that despite our encoders sharing the same architecture with the baseline,
by training our encoders through the Rank objective (Eq.~\ref{loss:rank}), our model is more effective at extracting disentangled pitch, rhythm, and timbre information.
It is worth mentioning that although both models use speaker ID to represent timbre, our model's disentanglement performance of timbre exceeds that of the baseline (e.g. the conversion rate results of Timbre-only). 
This may be due to our model's pitch, rhythm, or content representations containing less source's timbre information, resulting in less negative influence on the timbre conversion when converting the source speech's timbre to the target's.

% we apply pitch-only, rhythm-only, timbre-only, pitch+rhythm, pitch+timbre, rhythm+timbre, pitch+rhythm+timbre Conversion rate of these seven aspects on 20 utterance pairs, which are unseen during training that are perceptually distinct and present the results to five subjects. The test method of conversion rate is as follows: 
%  Figure 3 shows the conversion rates of different types of conversions. The conversion rate of the model in the corresponding aspects is high, while low otherwise. For example, the timbre-only conversion has a high timbre conversion rate but low pitch and rhythm conversion rates. We do not use the random resampling operation of baseline, but use the rank score method to decouple. The conversion rate results of seven aspects show that our model has relatively high conversion rate in each aspect and exceeds the baseline, indicating that only using the data enhancement method can achieve the decoupling effect.

\subsection{Naturalness MOS}

\begin{figure}[t]
  \centering
  \includegraphics[width=0.9\linewidth]{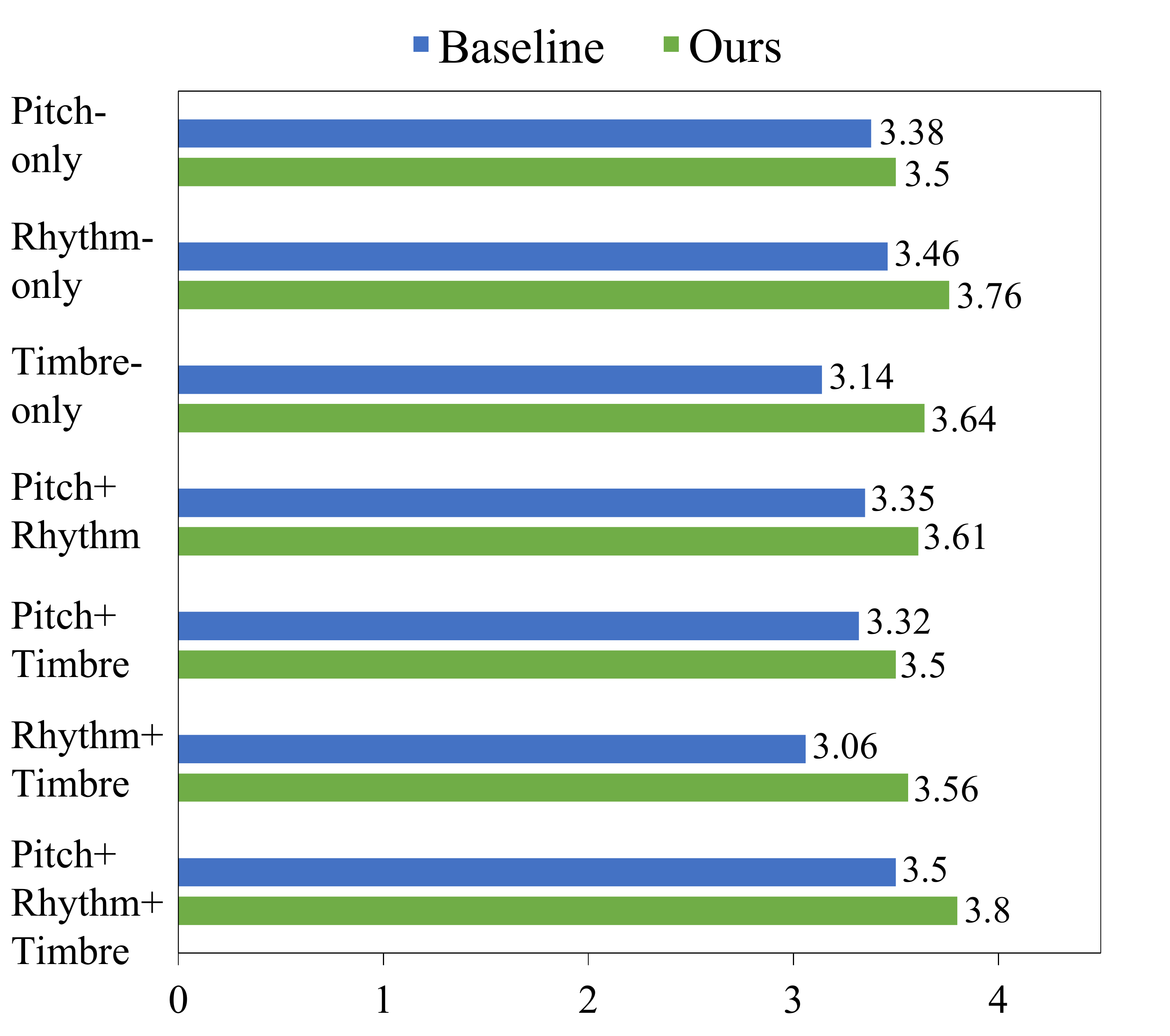}
  \caption{Naturalness MOS evaluation}
  \label{MOS}
  \vspace{-0.3cm}
\end{figure}

We then conduct Mean Opinion Score (MOS) to evaluate the naturalness of generated speech. 
We ask the subjects to score the speech's naturalness on a scale of 1 (Bad) to 5 (Excellent). The results are shown in Fig.~\ref{MOS}. 
All 7 combinations of different speech components have been tested.
As we can see, our model largely outperforms the baseline and can produce more natural speech.
High naturalness MOS scores indicate that the pitch, rhythm or timbre from the target speech is less likely to damage the naturalness of the generated speech.

% The conversion rate test is similar to the mean opinion score (MOS) similarity test. Therefore, in terms of another subjective evaluation, we conduct the MOS naturalness test. The total score is 5 points. the score of 1 indicates that the converted speech sounds completely unnatural, and the score of 5 indicates that the converted speech is close to the human voice.As shown in Figure 4, the naturalness scores of Pitch only, Pitch+Timbre are lower than other aspects, and the scores of Rhythm only is relatively high, which shows that our method can decouple the rhythm well. Although the model effect is better than baseline in terms of decoupling pitch, it still needs to be improved.

\subsection{Character Error Rate}

We continue to carry out Character Error Rate (CER) experiments. 
A pre-trained ASR model \cite{Gulati2020ConformerCT} is applied to produce CER.
CER quantifies the linguistic difference between original and converted speech to detect the ability of content preservation by models.

The results are shown in Tab.~\ref{cer}. As we can see, the error rate of our model is less than the baseline for almost all combinations, 
indicating that our model can preserve the content information better.
Such results demonstrate the effectiveness of the infoNCE loss (Eq.~\ref{loss:infonce}) we applied to force the content encoder to extract disentangled content representations. 
It is worth mentioning that the CER results are relatively high when converting the pitch component, possibly due to length mismatch between the source and target speech. 
If the target speech is longer than the source, conflicts between the target's long pitch and the original short rhythm or content information might occur, introducing noise into the generated speech.
% It is worth mentioning that the CER results are relatively high when we convert the pitch component. 
% This might be caused by the length mismatch between the source and target speech. 
% For instance, when the length of the target speech is longer than that of the source, 
% converting the source's pitch might result in conflicts between the target's long pitch and the original short rhythm or content information.
% This could introduce noise into the generated speech.

\subsection{Pearson Correlation Coefficient on Pitch}
We further verify the effectiveness of the pitch representation by measuring the Pearson Correlation Coefficient (PCC) \cite{cohen2009pearson} of logF0.
PCC measures the pitch correlation between two speech samples.
The larger the value, the more similar the pitch patterns are between the two samples.

The results are shown in Tab.~\ref{PCC}.
It is important to note that only two samples of the same length can be used to compute PCC. Thus, we split the results into two parts. 
For Pitch-only and Pitch+Timbre combinations, the converted and source speech have the same length. 
We expect a lower pitch correlation between converted and source speech due to pitch conversion. The PCC results of our model are lower than the baseline for these two combinations, indicating more effective pitch conversion.
Conversely, when converting the rhythm simultaneously, the length will be the same as the target, and we expect a higher PCC due to pitch transfer from the target speech. 
As we can see, Pitch+Rhythm and Pitch+Rhythm+Timbre combinations exhibit higher correlations than the baseline. 
In conclusion, our model can produce more disentangled and effective pitch representations.

% Pitch-only, Rhythm-only, Timbre-only, Pitch+Rhythm, Pitch+Timbre, Rhythm+Timbre, Pitch+Rhythm+Timbre

\begin{table}[t]
  \small
  \caption{CER ( \% )}
  \label{cer}
  \centering
    \begin{tabular}{ c | c | c}
    \toprule
    Combination                     & Baseline   & Ours             \\ 
    \midrule
    Pith-only    & 64.5     & \textbf{62.3} \\
    Rhythm-only    & 31.8     & \textbf{22.7} \\
    Timbre-only    & 50.1     & \textbf{46.0} \\
    Pitch+Rhythm    &\textbf{34.8}     & 36.2 \\
    Pitch+Timbre    & 61.8     & \textbf{60.0} \\
    Rhythm+Timbre    & 27.6     & \textbf{20.5} \\
    Pitch+Rhythm+Timbre    & 30.5     & \textbf{23.6} \\

    \bottomrule
    \end{tabular}
    % \vspace{-0.3cm}
\end{table}

% \begin{table}[t]
%   \small
%   \caption{Impact of pitch representation}
%   \label{PCC}
%   \centering
%     \begin{tabular}{  c | c | c | c | c }
%     \toprule
%         \multirow{2}{*}{Model}     & \multicolumn{2}{c|}{Source}   &\multicolumn{2}{c}{Target}          
%         \\ 
%         \cline{2-5}
%              & Baseline    & Ours & Baseline & Ours
%         \\
%         \cline{1-5}
%         P   & 0.45  & 0.41  &         & 
%         \\
%         % \cline{1-5}
%         PT   & 0.42  & 0.40  &         & 
%         \\
%         \cline{1-5}
%         RP  &        &       & 0.41   & 0.49
%         \\
%         % \cline{1-5}
%         RPT  &       &       & 0.53   & 0.69
%         \\

%     \bottomrule
%     \end{tabular}

% \end{table}

\begin{table}[t]
  \small
  \caption{Impact of pitch representation}
  \label{PCC}
  \centering
    \begin{tabular}{ c | c | c }
    \toprule
        Combination     & Baseline   & Ours          \\ 
        \midrule
        \multicolumn{3}{l}{PCC between the source and the converted} \\
        \multicolumn{3}{l}{(the lower the better)} \\
        \midrule
        Pitch-only      & 0.45   & \textbf{0.41}              \\ 
        Pitch+Timbre      & 0.42   & \textbf{0.40}              \\ 
        \midrule
        \multicolumn{3}{l}{PCC between the target and the converted} \\
        \multicolumn{3}{l}{(the higher the better)} \\
        \midrule
        Pitch+Rhythm      & 0.41   & \textbf{0.49}              \\ 
        Pitch+Rhythm+Timbre      & 0.53   & \textbf{0.69}              \\

    \bottomrule
    \end{tabular}
    \vspace{-0.3cm}
\end{table}

\section{Conclusions}

In this paper, we propose a VC model that is able to automatically disentangle the speech into four components without manually feeding various hand-crafted features or laborious bottleneck fine-tuning.
The disentanglement is derived from the Rank Module, which relies only on two augmentation transformations.
The subjective and objective evaluation shows that compared to the baseline, our model can achieve a better disentanglement performance and produce more natural speech.

\section{Acknowledgement}

This work was supported in part by the National Natural Science Foundation of China [grant number 61771196].

\bibliographystyle{IEEEtran}
\bibliography{mybib}

\end{document}